\documentclass[aps,twocolumn,floatfix,superscriptaddress]{revtex4}
\usepackage[version=3]{mhchem} 
\usepackage{graphicx}
\usepackage{dcolumn}
\usepackage{bm}
\usepackage{amsmath}
\usepackage{hyperref}
\usepackage{siunitx}  
\usepackage{amsfonts}
\usepackage{hhline}
\usepackage{MnSymbol}
\usepackage{float}
\usepackage{rotating}
\usepackage{color}

\newcommand{\ket}[1]{|{#1}\rangle} 

\newcommand{\ud}[1]{{#1^{\dagger}}}

\begin{document}
\flushbottom
\title{Spanning the full Poincar\'e sphere with polariton Rabi
oscillations}

\author{D.~Colas}
\affiliation{Departamento de F\'isica Te\'orica de la Materia Condensada and Condensed Matter Physics Center (IFIMAC), Universidad Aut\'onoma de Madrid, E-28049, Spain}

\author{L.~Dominici}
\email{lorenzo.dominici@gmail.com}
\affiliation{NNL, Istituto Nanoscienze-CNR, Via Arnesano, 73100 Lecce, Italy}
\affiliation{Istituto Italiano di Tecnologia, IIT-Lecce, Via Barsanti, 73010 Lecce, Italy}

\author{S.~Donati}
\affiliation{NNL, Istituto Nanoscienze-CNR, Via Arnesano, 73100 Lecce, Italy}
\affiliation{Istituto Italiano di Tecnologia, IIT-Lecce, Via Barsanti, 73010 Lecce, Italy}
\affiliation{Universit\'{a} del Salento, Via Arnesano, 73100 Lecce, Italy}

\author{A.A.~Pervishko}
\affiliation{Division of Physics and Applied Physics, Nanyang Technological University 637371, Singapore}

\author{T.C.H.~Liew}
\affiliation{Division of Physics and Applied Physics, Nanyang Technological University 637371, Singapore}

\author{I.A.~Shelykh}
\affiliation{Division of Physics and Applied Physics, Nanyang Technological University 637371, Singapore}
\affiliation{Science Institute, University of Iceland, Dunhagi 3, IS- 107, Reykjavik, Iceland}

\author{D.~Ballarini}
\affiliation{NNL, Istituto Nanoscienze-CNR, Via Arnesano, 73100 Lecce, Italy}

\author{M.~de~Giorgi}
\affiliation{NNL, Istituto Nanoscienze-CNR, Via Arnesano, 73100 Lecce, Italy}

\author{A.~Bramati}
\affiliation{Laboratoire Kastler Brossel, UPMC-Paris 6, ENS et CNRS, 75005 Paris, France}

\author{G.~Gigli}
\affiliation{NNL, Istituto Nanoscienze-CNR, Via Arnesano, 73100 Lecce, Italy}
\affiliation{Universit\'{a} del Salento, Via Arnesano, 73100 Lecce, Italy}

\author{E.~del~Valle}
\affiliation{Departamento de F\'isica Te\'orica de la Materia Condensada and Condensed Matter Physics Center (IFIMAC), Universidad Aut\'onoma de Madrid, E-28049, Spain}

\author{F.P.~Laussy}
\affiliation{Departamento de F\'isica Te\'orica de la Materia Condensada and Condensed Matter Physics Center (IFIMAC), Universidad Aut\'onoma de Madrid, E-28049, Spain}
\affiliation{Russian Quantum Center, Novaya 100, 143025 Skolkovo, Moscow Region, Russia}

\author{A.~V.~Kavokin}
\affiliation{Russian Quantum Center, Novaya 100, 143025 Skolkovo, Moscow Region, Russia}
\affiliation{CNR-SPIN, Tor Vergata, viale del Politechnico 1, I-00133 Rome, Italy}

\author{D.~Sanvitto}
\affiliation{NNL, Istituto Nanoscienze-CNR, Via Arnesano, 73100 Lecce, Italy}
\affiliation{Istituto Italiano di Tecnologia, IIT-Lecce, Via Barsanti, 73010 Lecce, Italy}

\begin{abstract}
  We propose theoretically and demonstrate experimentally a generation
  of light pulses whose polarization varies temporally to cover selected areas 
  of the Poincar\'e sphere with tunable swirling speed and total duration
  (\SI{1}{\pico\second} and \SI{10}{\pico\second} respectively in
  our implementation). The effect relies on the Rabi oscillations of
  two polarized fields in the strong coupling regime, excited
  by two counter-polarized and delayed pulses.  
  The interferences of the oscillating fields result in the
  precession of the Stokes vector of the emitted light while polariton
  lifetime imbalance results in its drift from a circle on the sphere
  of controllable radius to a single point at long times. The
  positioning of the initial and final states allows to engineer the
  type of polarization spanning, including a full sweeping of the
  Poincar\'e sphere. The universality and simplicity of the scheme
  should allow for the deployment of time varying polarization fields
  at a technologically exploitable level.
\end{abstract}

\date{\today} \maketitle

\section*{Introduction}

A new dimension has been literally opened for the control and
manipulation of light with ``\emph{polarization
  shaping}''~\cite{brixner01a,sato13a}. This makes the most out of the
vectorial nature of light by determining its time evolution not only
in phase and amplitude but also in its state of polarization. Since
the interaction of light and matter is polarization sensitive, the
control of this additional degree of freedom has allowed to outbeat
the performances of light in most of its usual applications, such as
photon ionization~\cite{brixner04a}, sub-wavelength
localization~\cite{aeschlimann07a} or timing with zeptosecond
precision~\cite{kohler11a}.  Proposals abound as to its future
applications in both a classical and quantum context~\cite{brif10a}.
Beyond the extension of the concept of pulse shaping to encompass
polarization, there has been as well increasing demand for
time-independent but spatially varying polarization
beams~\cite{brown10a} such as cylindrical vector
beams~\cite{youngworth00a}. When providing all the states of
polarization to realize so-called ``\emph{full Poincar\'e
  beams}''~\cite{beckley10a}, these also demonstrate advantages in
similar endeavors, such as boosted scattering or sub-wavelength
localization~\cite{beckley10a}. They also allow direct industrial
applications in laser micro-processing, such as improving the
efficiency and quality of processes like drilling holes for
fuel-injection nozzles~\cite{nolte99a}, processing of silicon
wafers~\cite{meier07a} or the machining of medical stent
devices~\cite{breitling_book04a}.  A new chapter of optics with
fundamental as well as applied benefits has therefore been started
with the availability of beams with a nontrivial dynamics of
polarization.

Here we show that semiconductor microcavities~\cite{kavokin_book11a}
in the strong exciton-photon coupling regime~\cite{weisbuch92a} offer
a convenient and powerful platform to bring together these two twists
on polarized light, by providing full Poincar\'e beams in time. This
is implemented in a largely self-contained integrated device of
micrometer size which does not rely on extrinsic processing of the
signal. As such, this considerably improves on the unwieldy complexity
of the setups required to generate polarization shaping and full
Poincar\'e beams independently, that both come with their respective
limitations.  Indeed, since conventional solid state lasers typically
emit light with a fixed linear polarization~\cite{baranov_book13a},
and optical nano-antennas also generally radiate a fixed polarization
determined by their geometrical
structure~\cite{rodriguez13a,banafsheh13a}, the synthesizing of a
desired polarization in space~\cite{lerman10a} or
time~\cite{brixner01a} is an involved process. Time varying
polarization is particularly demanding as it requires a combination of
liquid crystals, spatial light modulators, interferometers and
computer ressources with construction algorithms as well as state of
the art pulse-shaping techniques, that all together make up a complex
setup and impose some restrictions, such as the duration of the
pulse~\cite{polachek06a}.  The dynamics in our experiment takes place
in a femtosecond timescale, as in many pulse shaping counteparts, but
unlike these cases, it has no intrinsic restriction to a particular
timescale and turning to other systems with Rabi frequencies of
different magnitudes, the same dynamics can be realized with today's
technology in timeranges that run from attoseconds (with
plexitons~\cite{schlather13a,vasa13a}) to milliseconds (with
nanomechanical oscillators~\cite{faust13a}).  Our scheme also offers a
tunable and undemanding control of which area to span on the
Poincar\'e sphere and can be realized with any polarized fields that
supports strong coupling, not only light.  These results should allow
to disseminate the usage of time-dependent polarization beams in a
wide variety of platforms, including its extension to the quantum
regime by powering the mechanism with quantum Rabi oscillations.
\begin{figure}[t!]
  \includegraphics[width=.9\linewidth]{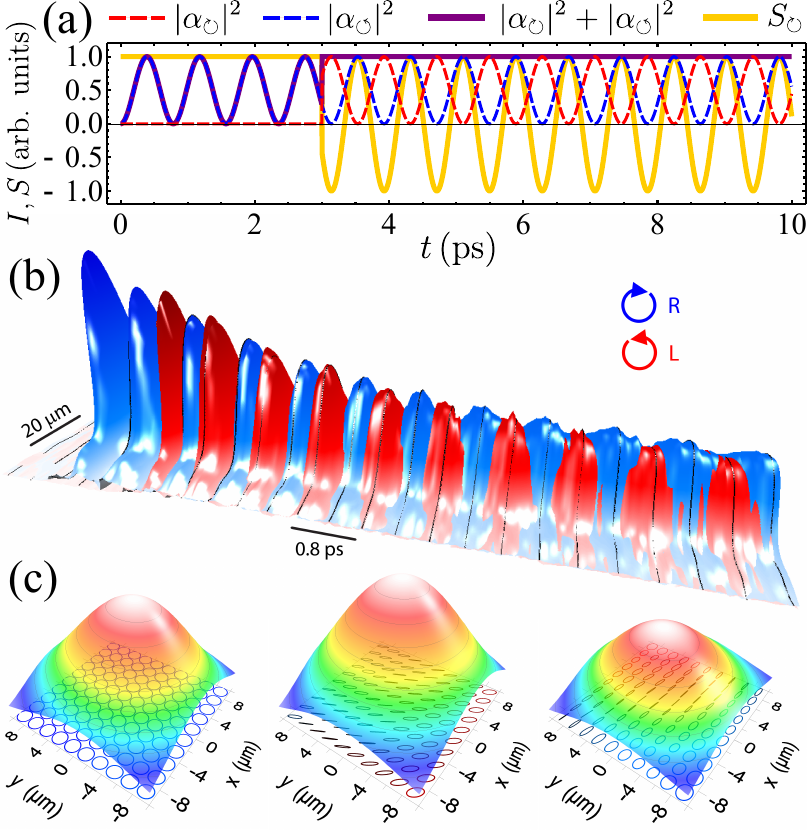}
  \caption{\textbf{Rabi oscillations of polarized beams.} (a) Rabi
    oscillations (theory) in i) one polarization only (till
    $t\approx\SI{3}{\pico\second}$), resulting in an oscillating
    intensity (purple line) of constant polarization (yellow line),
    and ii) in two polarizations, reversing the pattern of
    oscillations. (b) Experimental observation of the effect through
    the cavity field (along the diameter of the Gaussian spot of
    \SI{20}{\micro\meter} over a \SI{10}{\pico\second} duration) with
    left (red) and right (blue) circular polarization as a false color
    plot. In the experiment, the leakage of photons results in an
    exponential decay of the signal.  See also Supplementary Movies S1
    and S2.  (c) Spatial distributions of the density and polarization
    at \SI{200}{\femto\second} time intervals during one of the
    initial cycles after the second pulse arrival.  As can be seen
    while the emitted intensity remains basically constant (height
    scale in c) the resulting polarization is strongly affected by the
    Rabi oscillations (mutual oscillations in b and polarization map
    in c).  The polarization can also be made homogeneous or not
    spatially.}
  \label{fig:1}
\end{figure}
\section*{Principle of the mechanism}

The effect is based on the superposition of coherent states of
different polarizations in the regime of Rabi
oscillations~\cite{norris94a}. In some basis, say left
($\circlearrowleft$) and right ($\circlearrowright$) circular
polarization, the state of the system is described by four complex
amplitudes $\alpha_p$ and~$\beta_p$ with $p=\circlearrowleft$,
$\circlearrowright$ the state of polarization of the coherent
photon~$\alpha$ and exciton~$\beta$ fields, with wavefunction
$\ket{\psi(t)}=\ket{\alpha_\circlearrowleft(t),\beta_\circlearrowleft(t)}\ket{\alpha_\circlearrowright(t),\beta_\circlearrowright(t)}$,
where we have written the coupled exciton-photon fields in the same
ket vectors.  Since there is no coupling between the polarizations in
the linear regime of our experiment, each component evolves
independently with the second quantized Hamiltonian $\sum_p
g(\ud{a_p}b_p+a_p\ud{b_p})$ for the ladder operators $a_p$ (for the
photon) and~$b_p$ (exciton), at resonance and in the rotating
frame. Leaving aside for a moment the pulsed excitation to simply
consider the initial state, the solution reads~\cite{laussy09a}:
\begin{equation} 
  \label{eq:TueNov18161354CET2014} 
  \alpha_p(t)=\alpha_p(0)\cos(gt+\phi_{p})-i\beta_p(0)\sin(gt+\phi_p)\,,
\end{equation} 
The exciton solution reads similarly but it is not explicitly needed
since we are concerned in the time dynamics of the optical field only,
obtained by tracing over the excitons
$\ket{\psi_\alpha(t)}=\ket{\alpha_\circlearrowleft(t)}\ket{\alpha_\circlearrowright(t)}$
for
$\ket{\alpha_p}=\exp(-|\alpha_p|^2/2)\sum_{k=0}^\infty\alpha_p^k\ket{k}/\sqrt{k!}$
the coherent state of cavity photons with complex amplitude~$\alpha_p$
in the polarization~$p$. In the following, we will call
$\phi_\alpha=\phi_\circlearrowright-\phi_\circlearrowleft$ the
relative optical phase between the two photon fields at $t=0$, and
$\phi_\beta$ that between the exciton fields, which will play an
important role and which, in the experiment will be controlled by the
time delay between the exciting pulses.  From
Eq.~(\ref{eq:TueNov18161354CET2014}), it is apparent that the exciton
field is needed only as an initial condition. The state can be written
in any of the familiar forms to represent polarization, such as Stokes
or Jones vectors. In terms of the latter, the state of polarization
reads:
\begin{equation}
  \label{eq:10}
  \begin{pmatrix}
   \alpha_\leftrightarrow(t)\\
   \alpha_\updownarrow(t)
  \end{pmatrix}
  =
  \frac{\alpha_\circlearrowleft(t)}{\sqrt{2}}
  \begin{pmatrix}
    1\\i
  \end{pmatrix}
  +
  \frac{\alpha_\circlearrowright(t)}{\sqrt{2}}
  \begin{pmatrix}
    1\\-i
  \end{pmatrix}\,,
\end{equation}
with~$\alpha_{\leftrightarrow,\updownarrow}$ the electromagnetic field
components of the light in the linear polarization basis, emitted by
the cavity in the rotating frame.  One can then straightforwardly
obtain the intensity $\langle\ud{a_p}a_p\rangle$ and degree of
polarization
$S_p=(\langle\ud{a_p}a_p\rangle-\langle\ud{a_q}a_q\rangle)/(\langle\ud{a_p}a_p\rangle+\langle\ud{a_q}a_q\rangle)$
in any basis through Jones calculus or, equivalently, the
transformations
$a_\leftrightarrow=(a_\circlearrowleft+a_\circlearrowright)/\sqrt{2}$,
$a_\updownarrow=i(
a_\circlearrowleft-a_\circlearrowright)/\sqrt{2}$ and
$a_{\neswarrow/\nwsearrow}=((1\pm i)a_\circlearrowleft + (1\mp
i)a_\circlearrowright)/2$.

\begin{figure}[t!]
  \includegraphics[width=\linewidth]{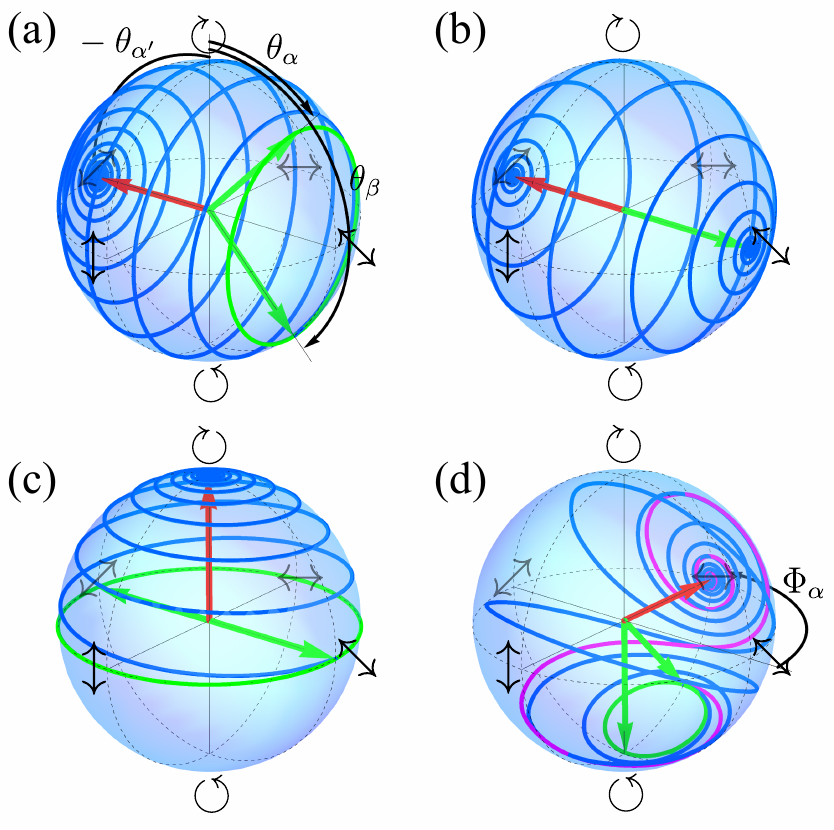}
  \caption{ \textbf{Dynamics of polarization on the Poincar\'e sphere
      (theory).} The green arrows, defined by the
    angles~$\theta_\alpha$ and~$\theta_\beta$, fix the circle of
    polarization in absence of decay ($\gamma=0$) by intersecting the
    meridian of azimuthal angle~$\phi_\alpha$. The red arrow, defined
    by the angle~$\theta_\alpha'$, fixes the point of long-time
    polarization. In presence of decay, $\gamma\neq0$, the trajectory
    of the polarization, in blue, drifts from the green circle to the
    red final point. (a) Span of the Poincar\'e sphere excluding a
    spherical cap of antidiagonal polarization, by setting
    $R_{\alpha}=0.41$, $R_{\beta}=2.41$ and
    $R_{\alpha}^{\prime}=-1$. (b) Span of the full Poincar\'e sphere
    from the antidiagonal to the diagonal pole, by setting
    $R_{\alpha}=1-\epsilon$, $R_{\beta}=1+\epsilon$ and
    $R_{\alpha}^{\prime}=-1$ for $\epsilon \rightarrow 0$. (c) Span of
    the northern hemisphere of the Poincar\'e sphere in circular
    polarization, by setting $R_{\alpha}=1-\epsilon$, $R_{\beta}=-1$
    and $R_{\alpha}^{\prime}\rightarrow \infty$ with $\epsilon
    \rightarrow 0$. (d) Distorted spanning of the sphere by choosing
    close initial and final points, by setting $R_{\alpha}=0$,
    $R_{\beta}=-2/3$, $R_{\alpha}^{\prime}=1$ and $\Phi_\alpha=\pi/2$.
    Parameters common to all cases: blue trajectories with $\gamma=1$
    and the purple trajectory with $\gamma=3$.}
  \label{fig:2}
\end{figure}

Figure~\ref{fig:1}(a) shows a simple illustration of this polarized
Rabi dynamics in the left-right circular polarization basis. The
intermittent transfer of light to the exciton field through Rabi
oscillations results in a temporary switch-off of the cavity emission
in the corresponding polarization. By synchronizing them so that the
cavity always emits in some polarization, one obtains a signal of
constant intensity but of oscillating polarization defined by their
superposition, which describes a circle on the Poincar\'e sphere
during one period of oscillations (shown as a green trace in
Fig.~\ref{fig:2}). The period, given by the polariton splitting, is
roughly equal to \SI{1}{\pico\second} in our case. The circle is
defined on the sphere, in a given basis (we will work in the circular
one), by two couples of angles
$(\theta_{\alpha,\beta},\Phi_{\alpha,\beta})$, defined by the ratios
of polarization
$R_{\alpha}=\alpha_\circlearrowleft/\alpha_\circlearrowright$ and
$R_{\beta}=\beta_\circlearrowleft/\beta_\circlearrowright$ of the
photon~$\alpha$ and exciton~$\beta$ fields at $t=0$, respectively. The
relation follows straightforwardly by geometric construction as:
\begin{subequations} 
 \label{eq:miénov26153758CET2014}
  \begin{align} 
	&\theta_\xi =2\arccos(1/\sqrt{1+|R_\xi|^2})\,,\label{eq:subeq1}\\
	&\Phi_\xi =\phi_\xi+\arg R_\xi\,,\label{eq:subeq2}
  \end{align} 
\end{subequations} 
for~$\xi=\alpha$, $\beta$. This is illustrated in
Fig.~\ref{fig:2}(a), where the case $\Phi_\xi=0$,
$\theta_\alpha=\pi/4$ and $\theta_\beta=3\pi/4$ is shown.  In the
particular case where $\theta_\alpha=\theta_\beta$ and $\Phi_\alpha=\Phi_\beta$, the circle reduces
to a point, i.e., to a constant polarization. This case corresponds to
choices of~$\alpha_p$ and~$\beta_p$ that define a polariton, i.e., an
eigenstate of the system with no temporal dynamics. The polarization
of light is then fixed to that of the corresponding polariton. 

\section*{Spanning the sphere thanks to polariton features}

We can now take advantage of a feature that is usually regarded as a
shortcoming of microcavity polaritons, but that in our case will turn
the simple effect just proposed into a mechanism that powers a new
type of light. There is a significant lifetime imbalance of the two
types of polaritons~\cite{dominici14a}, with the upper polariton
$\ket{\mathrm{U_p}}=\ket{\alpha_p,\beta_p}$---where
$\alpha_p=\beta_p$---being much more short-lived as compared to the
lower polariton $\ket{\mathrm{L_q}}=\ket{\alpha_q,\beta_q}$---where
$\alpha_q=-\beta_q$---regardless of the polarizations~$p,q$.  The
upper polariton lifetime is typically of the order of
\SI{2}{\pico\second} while the lower polariton lifetime is of the
order of \SI{10}{\pico\second}.  These values can however be tuned by
orders of magnitude with the already existing
technology~\cite{steger13a}.  This results in time-dependent Rabi
oscillations that converge toward a monotonously decaying signal as
the population of the upper polaritons ``evaporates'' and only lower
polaritons remain.
\begin{figure}[t!]
  \includegraphics[width=\linewidth]{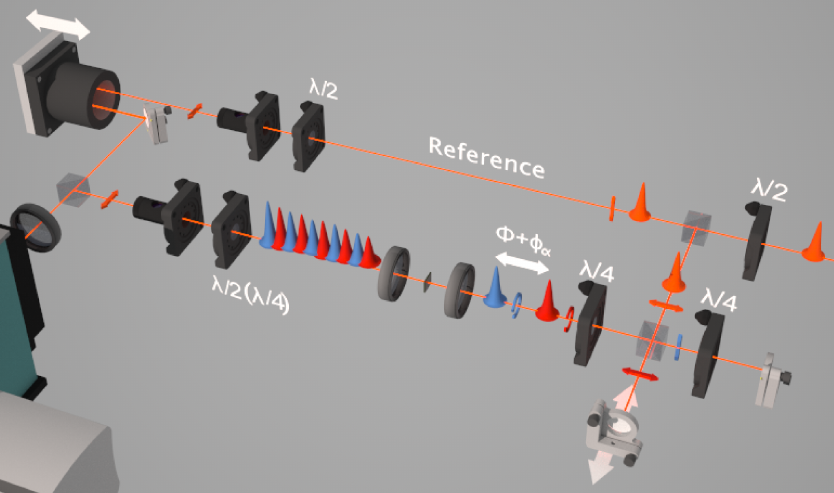}
  \caption{\textbf{Setup that implements the effect experimentally.}  The
effect is implemented and observed with a time-resolved digital
holography setup for counter-polarized double pulse experiments. The
fs pulses train is first split in a reference beam ($\updownarrow$)
and a signal beam ($\leftrightarrow$).  This latter is further divided
into twin pulses of equal $\leftrightarrow$ linear polarizations, of
which, only one, upon double passage onto a $\lambda/4$ plate, becomes
$\updownarrow$ linear.  After rejoining their paths, the twins are
made counter-circular (by use of a second $\lambda/4$ plate).  Their
mutual delay can be set on a Rabi period scale ($\Phi$) and wavelength
order ($\phi_{\alpha}$).  Sample emission is then filtered in
polarization and let to interfere with the reference on the camera,
before digital elaboration.}
  \label{fig:3}
\end{figure}
The overall dynamics of polarization is therefore that which starts by
describing the circle of the Rabi dynamics in absence of dissipation,
with a continuous drift towards the fixed point of polarization of the
lower polariton. The final state can be parametrized in the same way
with a couple of angles $(\theta'_{\alpha},\Phi'_{\alpha})$, by making
a rotation of the polariton basis. This introduces the parameters
$\alpha_p'=(\alpha_p-\beta_p)/\sqrt{2}$ and
$R_\alpha'=\alpha'_\circlearrowright/\alpha'_\circlearrowleft$ from
which one obtains the angles
$\theta'_\alpha=2\arccos(1/\sqrt{1+|R'_\alpha|^2)}$ and
$\Phi'_\alpha=\phi_\alpha+\arg R'_\alpha$
(cf.~Eqs.~(\ref{eq:miénov26153758CET2014})).

\begin{figure*}[t!]
  \includegraphics[width=\linewidth]{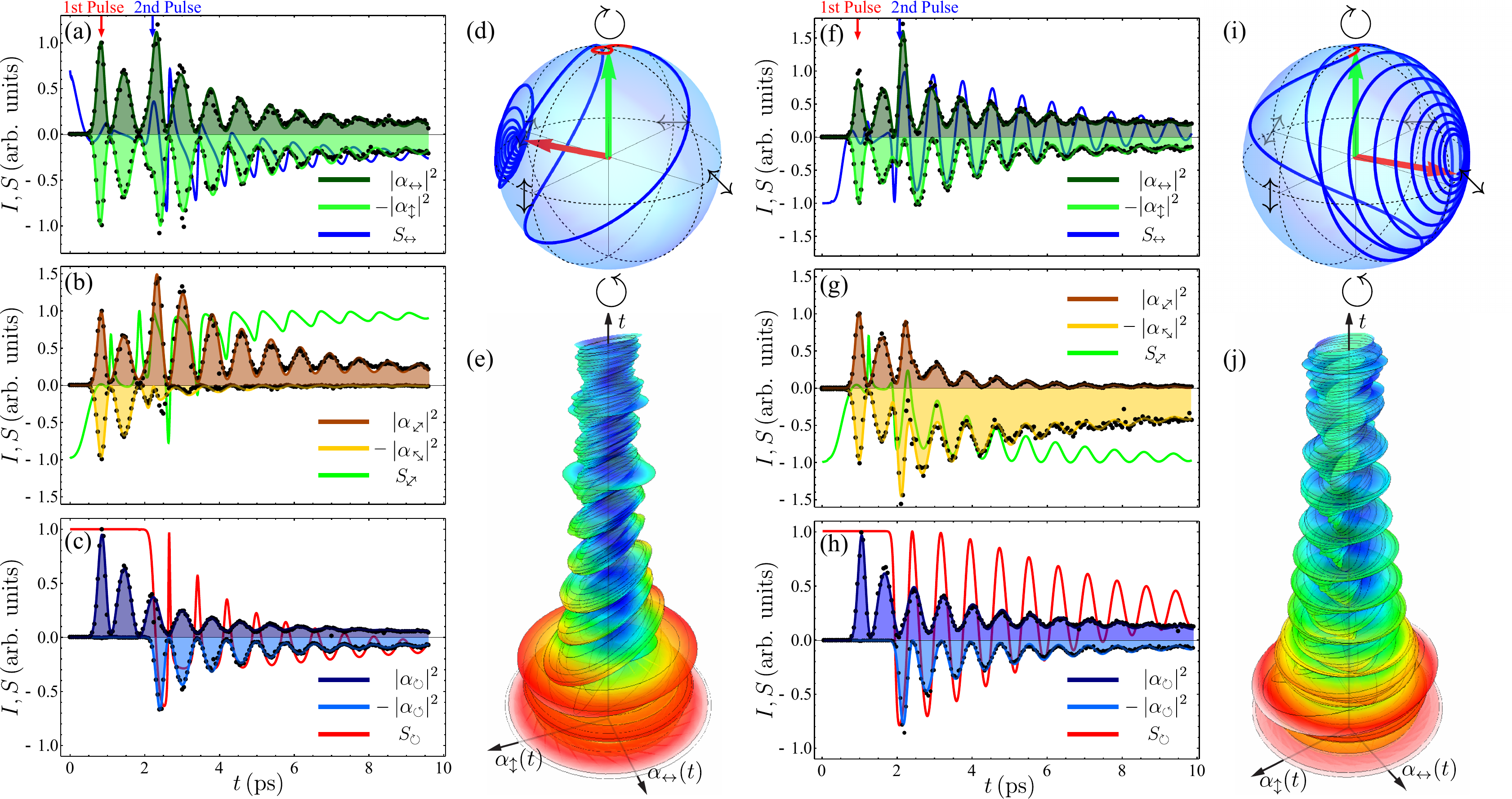}
  \caption{\textbf{Experimental Observation of the effect} Rabi
    in-phase experiment (a-e), i.e., the Rabi oscillations from the
    two different pulses start with the same phase, and Rabi
    antiphase experiment (f-j), i.e., the Rabi oscillations start
    with an opposite phase. The experimental data (a-c, f-h, black
    dots) is fitted by the theoretical model (solid lines), providing
    the amplitudes $|\alpha_{p,q}|^2$ and degrees of polarizations
    $S_p$. The dynamics of polarization can be displayed on the
    Poincar\'e sphere (d,i), demonstrating the rapid transition to the
    linear polarization in (d) and the spanning of the full hemisphere
    polarization in (i). (e,j) $2\textrm{D}+t$ representation of the
    field polarization, the radius and the color (from red to blue) of
    the ellipses are scaled on the instantaneous total intensity and
    polarization fields, respectively.}
  \label{fig:4}
\end{figure*}
By fixing with the effective polariton state of
Eq.~(\ref{eq:TueNov18161354CET2014}), on a meridian defined
by~$\phi_\alpha$, the initial and final states through the three
couples of angles~$(\theta_\alpha,\Phi_\alpha)$,
$(\theta_\beta,\Phi_\beta)$ (for the initial circle normal to the
meridian) and $(\theta'_{\alpha},\Phi'_{\alpha})$ (for the final point
on the meridian), one can thus span the Poincar\'e sphere of
polarization in essentially any desired way. The exact trajectory is
obtained by including decay in the dynamics of the coupled fields.
This is achieved by turning to a dissipative master
equation~\cite{delvalle_book10a} $\dot{\rho} = i\left [\rho, H \right
] + \mathcal{L}\rho$ for the four-fields density matrix, with
$\mathcal{L}\rho$ the Lindblad super-operator.
Assuming radiative decays of all fields (with their corresponding
decay rates~$\gamma_{a,b}$) as well as a mechanism dephasing the upper
polariton, that can be either radiative
decay~$\gamma_\mathrm{U}^{\mathrm{R}}$ or pure
dephasing~$\gamma_\mathrm{U}^\phi$, and incoherent pumping~$P_b$ from
the exciton reservoir, we are brought to a Liouvillian in the
form~\cite{dominici14a} $\mathcal{L}\rho
=\sum_{p=\circlearrowright,\circlearrowleft}\Big[
\frac{\gamma_{a}}{2}\mathcal{L}_{a_p} +
\frac{\gamma_{b}}{2}\mathcal{L}_{b_p} +
\frac{P_b}{2}\mathcal{L}_{\ud{b_p}}+
\frac{\gamma_{\mathrm{U}}^\mathrm{R}}{2}\mathcal{L}_{u_p}+
\frac{\gamma_{\mathrm{U}}^\phi}{2}\mathcal{L}_{\ud{u_p}u_p}\Big]\rho$,
with $\mathcal{L}_c \rho = 2 c\rho\ud{c} - \ud{c}c\rho - \rho\ud{c}c$
for the generic operator~$c$. 
This too can be solved in closed-form. For the observables of
interest, the solution remains fully defined by complex amplitudes:
\begin{multline} 
    \label{eq:juejun5211819CEST2014} 
    \alpha_p(t) = \Bigg[\alpha_p(0)\cosh(R t/4+i\phi_p)\\
    -\left(\frac{\beta_p(0) G + \alpha_p(0)
        \Gamma}{R}\right)\sinh(Rt/4+i\phi_p)\Bigg]\exp(-\gamma t/4)\,,
 \end{multline} 
where we have introduced: 
\begin{subequations} 
  \label{eq:juejul10095249CEST2014}
  \begin{align} 
    &\gamma= \gamma_a+\gamma_b+\gamma_\mathrm{U}-P_b\,,&
    \Gamma= P_b -\gamma_b+\gamma_a\,,\\
    &G = i4g +\gamma_\mathrm{U}\,,
    &R = \sqrt{G^2+\Gamma^2}\,,\\
    &\gamma_\mathrm{U}=\gamma_\mathrm{U}^\mathrm{R}+\gamma_\mathrm{U}^\phi\,.\label{eq:marnov18225748CET2014}
  \end{align} 
\end{subequations} 
Equation~(\ref{eq:marnov18225748CET2014}) states that it does not
matter which mechanism is responsible for the loss of the upper
polaritons, only the total decay rate enters the dynamics.

Some illustrative examples of this dynamics are shown in
Fig.~\ref{fig:2}. In all cases, green arrows point at the initial
states and the red one at the fixed point of the asymptotic final
state. The green circle shows the polarization cycle in absence of
dissipation and the blue trajectory the spanning of the Poincar\'e sphere for
the system parameters (given in the caption). Panel~(a) shows the
spanning excluding a spherical cap of antidiagonal
polarization. Panel~(b) includes it by reducing the circle to a single
point, thereby achieving a full spanning of the sphere. Panel~(c)
shows the spanning of a hemisphere only and from the
horizontal/diagonal basis equator towards the right-circular
polarization, again, covering a different area merely by tuning the
parameters~$\alpha_p$, $\beta_q$ and~$\phi_\alpha$. Panel~(d) shows
two interesting variations of this effect: first, by choosing close
points on the sphere, one can obtain a twisted trajectory, covering
different areas with different speeds and, therefore, opening the
possibility for emitters of pseudo-random polarization given the
uncertainty at which time the photon will be emitted along an
intricate path on the sphere.  Second, by varying the Rabi period,
which can be achieved by tuning the coupling strength, or the
decay rates, one can vary the number of loops around the
sphere.

\section*{Experimental implementation}

To implement these effects in the laboratory, we used a GaAs polariton
microcavity in the regime of Rabi oscillations described in previous
works~\cite{dominici14a,ballarini13a}. A sketch of our setup is shown
in Fig.~\ref{fig:3}, which is based on the principles of the
time-resolved digital holography~\cite{dominici14a}.  Two femtosecond
pulses with adjustable delay and polarization excite the system. The
pulses are initially linearly polarized and subsequently passed
through quarter wavelength plates to make them counter-circularly
polarized.  The energy spread of the pulses overlaps with both
polariton branches and thus triggers Rabi oscillations between
excitons and photons. After the 1st pulse, the lower and upper
polaritons have the same circular polarization.  After the 2nd pulse,
the dynamics of polarization is triggered according to the principle
of Eq.~(\ref{eq:juejun5211819CEST2014}). By weighting adequately the
two branches and by adjusting the optical phase between the two
pulses, we are able to realize various cases of interest predicted by
the theory.  Some typical experimental observations are shown in
Fig.~\ref{fig:4}, where we present the detected polarization in all
the bases, namely (a) $\leftrightarrow$/$\updownarrow$, (b)
$\nwsearrow$/$\neswarrow$ and (c)
$\circlearrowleft$/$\circlearrowright$. We show in each case the
intensity of light emitted in each component, $|\alpha_p|^2$ and the
corresponding degree of polarization $S_p$.  The points are
experimental data and the lines are theory fits with the model
presented above but supplemented with the dynamics of excitation by
femtosecond polarized pulses. While all polarizations are measured
experimentally, only two polarizations are needed by the theory to
obtain the other ones.  We have checked the consistency of the model
and the observation by fitting polarizations in all bases and by their
reconstruction from one basis only.  Experimentally, only the photonic
field is accessible, but the theory allows to reach the exciton field
as well. Therefore, we are able to reconstruct the polarization
dynamics, as shown in Fig.~\ref{fig:4} (d,i), and (e,j) with a 3D
(2D+time) representation of the polarization. The effective quantum
state can also be reconstructed~\cite{dominici14a}. The dynamics is
more clearly visualized on the Poincar\'e sphere. Figure~\ref{fig:4}
shows how the beam of light emitted by the microcavity provides an
ultra-fast sweeping of, in these cases, a hemisphere of the Poincar\'e
sphere, and with two speeds of relaxation similarly to the cases of
the theoretical model shown in Fig.~\ref{fig:2}(d).  In all cases, the
agreement between theory and experiment is excellent.

\section*{Extensions and applications}

In the literature, a variety of multiple polarized beams are
implemented by setting space profiles with different polarization.  A
noticeable example is that of radial (hedgehog) or azimuthal
polarization field states realizable, e.g., by use of Q-plate
devices~\cite{cardano12a}.  Here, we have used essentially spatially
homogeneous profiles in the experiments, as seen in the time-space
chart in Fig.~\ref{fig:1}(b,c), to focus on the time dynamics
instead. Nevertheless, a spatially dependent polarization could also
be combined with the temporal dynamics we have highlighted.  One of
the easiest space patternings would consist of sending the second
pulse with a slight angle of incidence (i.e., a $\Delta k_{\alpha}$)
with respect to the first one. In this case, while co-polarized beams
would give interference fringes of amplitude, Rabi-oscillating in time
(hence moving with a $g/\Delta k_{\alpha}$ velocity), in the case of
counter-polarized beams, all the dynamics discussed in the present
work could be obtained with an associated phase delay between the
fringes. Each fringe could be made time-oscillating in polarization
and with a phase offset with respect to each other, giving rise to a
flow of polarization waves with Rabi time period and settable space
period, the whole drifting towards the fixed polarization state of the
LP polariton. Such effects are beyond the scope of this work but give
a hint as to the rich patterns of polarization texture that are within
reach, when powered by the dynamics of polariton
fluids~\cite{carusotto13a}.

\section*{Acknowledgments}

 We acknowledge funding from the ERC Grant POLAFLOW, the IEF  project SQUIRREL (623708) and the support from IRSES project POLAPHEN.

\section*{Supplementary Material}

Supplementary Movie S1. The ultrafast imaging sequence of the 
opposing polarization densities in the experiment of Fig.~\ref{fig:1}(b,c). 
The map is over a \SI{30x30}{\micro\meter} area and along a 
\SI{10}{\pico\second} duration with time step of \SI{50}{\femto\second}.

Supplementary Movie S2. The dynamics in a time-space representation, 
with the amplitude of the two circular polarizations plotted vs a central diameter and time.


\end{document}